\documentclass[manuscript]{rmaa}

\usepackage{paralist}

\usepackage{psfrag,color}
\usepackage{longtable}

\usepackage[latin1]{inputenc}

\title{A sensitive search for methanol line emission toward evolved stars} 

\author{
  Jos\'e F. G\'omez,\altaffilmark{1} 
  Lucero Uscanga,\altaffilmark{2}
  Olga Su\'arez,\altaffilmark{3}
  J. Ricardo Rizzo,\altaffilmark{4}
  and Itziar de Gregorio-Monsalvo\altaffilmark{5,6}}

\altaffiltext{1}{Instituto de Astrof\'{\i}sica de Andaluc\'{\i}a,
  CSIC, Spain.}
\altaffiltext{2}{Institute of Astronomy, Astrophysics, Space
Applications and Remote Sensing, NOA, Greece}
\altaffiltext{3}{Laboratoire Lagrange, UNSA-OCA-CNRS, France.}
\altaffiltext{4}{Centro de Astrobiolog\'{\i}a, INTA-CSIC, Spain.}
\altaffiltext{5}{Joint ALMA Observatory, Chile}
\altaffiltext{6}{European Southern Observatory, Germany.}

\shortauthor{G\'omez et al.}
\shorttitle{Methanol in evolved stars}

\fulladdresses{
\item Itziar de Gregorio-Monsalvo: European Southern Observatory, Karl Schwarzschild Str 2, 85748 Garching bei M\"unchen, Germany (idegrego@alma.cl).
\item Jos\'e F. G\'omez: Instituto de Astrof\'{\i}sica de Andaluc\'{\i}a,
  CSIC, Apartado 3004, E-18080 Granada, Spain (jfg@iaa.es).
\item J. Ricardo Rizzo: Centro de Astrobiolog\'{\i}a, INTA-CSIC, Ctra. M-108, km. 4, E-28850 Torrej\'on de Ardoz, Spain (ricardo@cab.inta-csic.es).
\item Olga Su\'arez: Laboratoire Lagrange, UMR 7293, Universit\'e de Nice Sophia-Antipolis, CNRS, Observatoire de la C\^ote d'Azur, 06300 Nice, France (olga.suarez@oca.eu).
\item Lucero Uscanga: Institute of Astronomy, Astrophysics, Space
Applications and Remote Sensing, National Observatory of Athens, I. Metaxa \& V. Pavlou, P. Penteli, 15236 Athens, Greece (lucero@astro.noa.gr).}

\listofauthors{J. F. G\'omez, L. Uscanga, O. Su\'arez, J. R. Rizzo, \& I. de Gregorio-Monsalvo}
\indexauthor{G\'omez, J. F.}
\indexauthor{Uscanga, L.}
\indexauthor{Su\'arez, O.}
\indexauthor{Rizzo, J. R.}
\indexauthor{de Gregorio-Monsalvo, I.}

\abstract{We present a sensitive search for methanol line emission in evolved stars  at 1 cm, aiming to detect, for the first time, methanol masers in this type of objects. Our sample comprised post-AGB stars and young planetary nebulae (PNe), whose mass-loss processes and circumstellar structures resemble those of young stellar objects (YSOs), where methanol masers are detected. Class I masers were searched for in 73 objects, whereas Class II ones were searched in 16. No detection was obtained. The non-detection of Class I methanol masers indicated that methanol production in dust grains and/or the enhancement of its gas-phase abundance in the shocked regions of evolved objects are not as efficient as in YSOs. We suggest that relatively more evolved PNe might have a better probability of harboring Class II masers.}

\resumen{Presentamos una b\'usqueda muy sensible de emisi\'on de
  l\'{\i}neas de metanol a 1 cm en estrellas evolucionadas, 
  con el fin de detectar, por primera vez, m\'aseres de metanol
  en este tipo de objetos. Nuestra muestra incluye estrellas
  post-AGB y nebulosas planetarias (NP) j\'ovenes, cuyos procesos de
  p\'erdida de masa y sus estructuras circunestelares se asemejan a
  los de objetos estelares j\'ovenes (OEJ), en los que se detectan
  m\'aseres de metanol. Buscamos m\'aseres de Clase I en 73 objetos y
  de Clase II en 16. No encontramos ninguna detecci\'on. La ausencia
  de detecciones de m\'aseres de Clase I indica que la producci\'on de metanol en granos de polvo, o el incremento de su
  abundancia en las zonas de choque de objetos evolucionados no es tan
  eficiente como en OEJs. Proponemos que las NPs relativamente m\'as evolucionadas podr\'{\i}an tener una mayor probabilidad de detecci\'on de m\'aseres de Clase II.}

\addkeyword{ISM: molecules}
\addkeyword{masers}
\addkeyword{radio lines: stars}
\addkeyword{stars: AGB and post-AGB}

\begin{document}
\maketitle

\section{Introduction}
\label{sec:intro}

The methanol molecule is rich in transitions with frequencies in the radio regime (mm to cm wavelengths), of which more than 20 are known to emit as masers \citep{sob93}.
Methanol masers are found in many star-forming regions, usually associated with high-mass young stellar objects (YSOs) \citep[see, e.g.,][]{has90,men91a,pes05,pra08}, but they are also present around a few low-mass ones \citep{kal06,kal10}. However, there is no reported detection of methanol lines (either thermal or maser) toward evolved objects, despite several searches \citep{bac90,lat96a,lat96b,cha97}. A possible exception may be IRAS 19312+1950 \citep{deg04}, although the nature of the object and whether the methanol lines are really associated with it, or with a nearby molecular cloud is still unclear. Other prospective cases of methanol in evolved stars \citep{wal03,urq13} have been refuted by \citet{bre13}.

The non-detection of {\em thermal lines} of methanol ruled out chemistry models \citep{cha95,mar05} that invoked an injection of CH$_4$ from the inner to the outer circumstelar envelope, as a way to explain the high abundance of HCN observed in some oxygen-rich asymptotic giant branch (AGB) stars \citep{buj94,bie00}. These models predicted an enhancement of methanol abundance that should have resulted in flux densities of its thermal lines above the sensitivity limit of the observations.

Moreover, the absence of {\em methanol masers} in evolved stars suggests that the energy input and/or the physical conditions of density or temperature necessary to maintain the inversion of population, are not met in this type of objects, in contrast with YSOs. Alternatively, it is possible that the physical conditions are appropriate for maser pumping, but the methanol molecule is not abundant enough to produce any significant emission. This situation is significantly different from the case of other molecular species, such as SiO, OH, and H$_2$O, whose maser emission is detected toward both young and evolved objects \citep{eli92}. 

Searches for methanol in evolved stars have focused on AGB stars, which is understandable because they have dense circumstellar envelopes, and the detection rates of SiO, OH, and H$_2$O masers are much higher than in subsequent phases \citep{lew89,gom90}. 
However, the next evolutionary stages may provide more adequate targets. In post-AGB stars and early planetary nebulae (PNe) the structure of the circumstellar medium and the physical processes are similar (at least qualitatively) to those of YSOs \citep[see, e.g.,][]{shu87}. They often display collimated jets \citep{sah98,gom11}, circumstellar disks \citep{kwo00,buj05}, and thick envelopes surrounding the whole system \citep{ima09,ram09,riz13}.

Considering these similarities, we have carried out a search for methanol masers focused on post-AGB and young PN candidates, using two of the most sensitive single-dish radio telescopes in the world working at centimeter wavelengths. We have searched for emission in four different transitions, pertaining to the two different classes of methanol masers, commonly known as class I and II \citep{bat87,men91b}.

\section{Source Sample}

We have used two different samples in our observations. The Class II 
masers were searched for in some post-AGB stars and PNe from the catalogue of 
\citet{sua06} that fulfilled at least one of these criteria: (a) 
they had previous detection of either H$_2$O or OH masers, (b) they were classified as 
``transition objects'' in \citet{sua06}, or (c) they were PNe of low excitation. In addition, we 
included some post-AGB stars and PNe that were not included in \citet{sua06}, but that showed SiO masers 
\citep{nym98}.

The second sample used for the Class I maser search comprises most of the northern sources in \citet{ram09}. They are post-AGB stars and PN candidates with the IRAS color criteria of \citet{sua06}, and with signs of strong optical obscuration.

\section{Observations}
\label{sec:obs}

\subsection{Robledo de Chavela}

The Class II $2_1-3_0$ E transition of methanol (rest frequency = 19967.3961 MHz) was observed with the DSS-63 antenna, located in the Madrid Deep Space Communications Complex (MDSCC), near Robledo de Chavela, Spain. The antenna has a diameter of 70 m, providing a half-power beam width of 46" at this frequency. The observations were carried out  between September 2002 and June 2005, using two different backends depending on the observing dates. In 2002 and 2003 we used a 256-channel spectrometer covering a bandwidth of 10 MHz, which provided a velocity resolution of $\simeq 0.59$ km s$^{-1}$. In 2005 we used a 384-channel spectrometer covering a bandwidth of 16 MHz ($\simeq 216$ km s$^{-1}$ with $\simeq 0.63$ km s$^{-1}$ resolution).  Spectra were taken in position-switching mode with the 384-channel spectrometer and in frequency switching mode, with a switch of 5 MHz, when using the 256-channel one, thus providing in the latter case an effective velocity coverage of $\simeq 202$ km s$^{-1}$ (15 MHz). Only left-circular polarization was processed. The total integration time was typically 20 minutes per source in frequency-switching mode and 30 minutes (on + off) in position-switching mode 
The rms pointing accuracy was better than $10''$. The data reduction was performed using the CLASS package, which is part of the GILDAS software. Spectra were corrected of the elevation-dependent gain of the telescope and of atmospheric opacity.

\subsection{Green Bank Telescope}

We observed three class I methanol transitions of the J$_2-$J$_1$ E series: $2_{2}-2_{1}$, $3_{2}-3_{1}$, and 
$4_2-4_1$, with rest frequencies 24934.382, 24928.707, and 24933.468 MHz, respectively, using the Robert C. Byrd Green Bank Telescope (GBT) of the National Radio Astronomy Observatory, in March 2010. The 1.3 cm receiver comprised four beams, arranged in two pairs that could be tuned independently. We used one such pair, with a separation of $178.8''$ between the beams, to simultaneously observe an on- and off-source position in dual polarization. Antenna nodding between the two beams was used to subtract atmospheric and instrumental contributions. We selected two spectral windows, one comprising the three methanol lines (and centered at the frequency of the $4_{2}-4_{1}$ E transition), and the other one centered at the water line at 22 GHz (whose results will be presented in G\'omez et al. in preparation). The bandwidth of each spectral window was 50 MHz ($\simeq 601$  km s$^{-1}$ velocity coverage) sampled over 8192 channels ($\simeq 0.07$ km s$^{-1}$ velocity resolution). The
half-power beam width of the telescope was $30''$ at the frequency of the methanol lines.
The integration time per source was $\simeq 4$ minutes, and all of it was effectively on-source time, because of the use of the antenna-nodding with dual beams. The data reduction was carried out with the GBTidl package. Spectra were corrected of the elevation-dependent gain of the telescope and of atmospheric opacity. 
The rms pointing accuracy was better than $3''$. 

\section{Results and discussion}
\label{sec:results}

No emission of methanol was detected with either telescope. 
Tables \ref{tab:robledo} and 2 give the ($1\sigma$) rms noise level of each spectrum. 
We consider that our upper limits for detections are $\simeq 3\sigma$ levels. 

The non-detection of thermal emission from these transitions is not surprising, since an 
unreasonably high abundance of methanol would be required to obtain mensurable emission. 
We have calculated upper limits to the abundance of methanol with respect to hydrogen, 
following the formulation in \citet{cha97}, and assuming  physical parameters appropriate 
for circumstellar envelopes expelled in the AGB phase  \citep[see, e.g.,][]{cha97,mar05}: expansion velocity of 15 km s$^{-1}$, 
temperature of 30 K, mass loss rate of $10^{-5}$ M$_\odot$ yr$^{-1}$, and distance from 
the star to the peak of molecular emission of $10^{16}$ cm. The solar distance is 
unknown for most of these objects, but we have assumed 8 kpc in our calculations, 
since a significant fraction are in the direction of the Galactic bulge.
The more restrictive limit for the abundance is $< 0.03$, obtained with the 
$4_2-4_1$ transition. Such an extremely high limit obviously does not give any useful information. 
Even taking into account the large uncertainties for 
the assumed parameters, 
far more restrictive upper limits to the methanol abundance in evolved stars 
have been obtained elsewhere with millimeter methanol lines, on the order of $5\times 10^{-9}-6\times 10^{-7}$ \citep{cha97,mar05}.

However, more relevant here is the fact that no maser emission has been detected.  As mentioned above, we have included lines pertaining to both types of methanol transitions (Class I and II). Class II methanol masers have been found exclusively toward high-mass YSOs \citep{bre13}.
They tend to be closely associated with HII regions, and are believed to be excited by intense infrared radiation fields.  In the particular case of the  $2_1-3_0$ E transition we observed, its emission seems to be correlated with the flux density of the background radio continuum emission \citep{kri13}. Although young PNe in our sample also show free-free radio continuum emission, the non-detection of $2_1-3_0$ E masers may indicate that  the radiation field from the central star is not enough to create population inversion, that the radio continuum emission is too weak to ignite the maser, or that methanol abundance is too low. It is possible that a search for this line emission in more evolved PNe, whose radio continuum flux density is stronger, would have a higher probability of finding detections.

On the other hand, Class I methanol masers are thought to be collisionally pumped. They are found in the neighborhood of both low- and high-mass YSOs, but their locations are offset from those of Class II masers. It is possible that Class I masers arise in post-shock gas
in the lobes of bipolar outflows, where the abundance of methanol
is enhanced due to grain mantle evaporation \citep{pla90}, and the energy released  inverts the populations of the molecule. One might expect that Class I masers could also be present in some post-AGB star, since they eject collimated jets  in a similar way as the case of low-mass YSOs \citep[e.g.,][]{sah02}.  However, it is interesting that the conditions for water and OH maser pumping are met in both young and evolved objects, but it is not the case of methanol for the latter. In particular, 
we note that some post-AGB stars and young PNe show water maser emission, which in the case of "water fountains" trace jets with high velocities ($\ga 50$ km s$^{-1}$) and collimation degrees. Water masers can originate in J-shocks, involving high-velocities ($> 40$ km s$^{-1}$), in regions with densities $n\simeq 10^6-10^8$ cm$^{-3}$, but they are quenched at $n \ga 10^8$ cm$^{-3}$ \citep{hol13}. Interestingly, these density conditions are very similar to those for the $J_2-J_1$ E maser lines of methanol \citep{leu04} that we observed with the GBT. The fact that water masers are detected, but methanol ones are not, strongly suggests that, although the methanol energy levels could be actually inverted, the column density of the molecule in shocked regions of post-AGB stars and young PNe is not large enough to produce detectable emission. This indicates that the production of methanol molecules in dust grains \citep{gar06,bre13} and/or the enhancement of its gas-phase abundance in the shocked regions of evolved objects are not as efficient as in YSOs.

\acknowledgements{JFG, OS, and IdG acknowledge support from MICINN (Spain) AYA2011-30228-C03 grant (including FEDER funds). L.U. acknowledges support from
grant PE9-1160 of the Greek General Secretariat for Research and Technology
in the framework of the program Support of Postdoctoral Researchers. JRR acknowledges support from MICINN grants CSD2009-00038, AYA2009-07304, and AYA2012-32032.  The National Radio Astronomy Observatory is a facility of the National Science Foundation operated under cooperative agreement by Associated Universities, Inc. This work is partially based on observations done at MDSCC (Robledo), under the Host Country Radio Astronomy program. Part of the data in this paper was reduced using the GILDAS software package (http://www.iram.fr/IRAMFR/GILDAS).}

\newpage

\begin{table*}[!t]\centering
  \setlength{\tabnotewidth}{0.95\columnwidth}
  \tablecols{6}
  \setlength{\tabcolsep}{1.2\tabcolsep}
  \caption{Robledo observations} \label{tab:robledo}
 \begin{tabular}{lrrrrr}
    \toprule
IRAS name & RA(J2000) & Dec(J2000) & rms\tabnotemark{a}  & Date\tabnotemark{b} & $V_\circ$\tabnotemark{c}  \\  
       & (h:m:s)   & ($^\circ$:$'$:$''$)   & (Jy) &  (yyyy-mm-dd) & (km s$^{-1}$)\\
    \midrule
 06530$-$0213  & 06:55:32.07 & $-02$:17:30.1 & 0.03  & 2003-05-16 & +28.0  \\
 16559$-$2957  & 16:59:08.22 & $-30$:01:40.8 & 0.18  & 2003-05-16 & +56.0 \\  
 17086$-$2403   &  17:11:38.76 & $-24$:07:33.5&  0.04 & 2005-06-16 & $-$0.3 \\
 17291$-$2402   &  17:32:12.98 & $-24$:05:00.8 &  0.04 & 2005-06-16 & $-$0.3  \\
 17347$-$3139   & 17:38:01.28 & $-31$:40:58.1 &  0.07 & 2005-06-16 & $-$68.3  \\
 17395$-$0841  & 17:42:14.08 & $-08$:43:21.6 & 0.12  & 2003-05-16 & +85.0  \\   
 18061$-$2505   & 18:09:12.54 & $-25$:04:35.5 & 0.05 & 2005-06-17 &   +3.8  \\
 19016$-$2330   & 19:04:43.21 & $-23$:26:11.0 &  0.05 & 2005-06-17 & $-$0.2 \\
 19154$+$0809  &  19:17:50.68 & +08:15:06.0 & 0.10    & 2002-10-11 &  0.0  \\
  19255+2123     &  19:27:43.99 & +21:30:03.0 & 0.05    & 2002-09-05 & +25.0  \\
  \multicolumn{1}{c}{''}  &  \multicolumn{1}{c}{''} &   \multicolumn{1}{c}{''}    & 0.25     & 2002-10-11 & +25.0   \\
 19590$-$1249   & 20:01:49.77 & $-12$:41:17.2 &  0.04 & 2005-06-17 &   84.8  \\
 20406$+$2953   & 20:42:45.95 & +30:04:06.0  & 0.12     & 2002-10-11 & +15.0   \\
 20462$+$3416   &  20:48:16.64 & +34:27:24.2 & 0.08    & 2002-09-05 &  0.0  \\
 21546$+$4721  &  21:56:33.03 & +47:36:13.0 & 0.07    & 2002-09-05 &  0.0  \\
 22023$+$5249  &  22:04:12.24 &  +53:03:59.9 & 0.05    & 2002-09-05 &  0.0   \\
 22036$+$5306   & 22:05:30.63 & +53:21:32.6  & 0.13     & 2002-10-11 & $-40.0$  \\
    \bottomrule
    \tabnotetext{a}{One-sigma rms noise level of the spectra, after correction by elevation-dependent gain and atmospheric opacity}
    \tabnotetext{b}{Date of observation}
    \tabnotetext{c}{Central LSR velocity of the spectra}
  \end{tabular}
\end{table*}

\newpage

\setlength{\tabnotewidth}{0.95\columnwidth}
\tablecols{5}
\tabcaption{GBT observations\tabnotemark{a}\label{tab:gbt}} 
\setlength{\doublerulesep}{0pt}
\setlength{\overfullrule}{1em}
\begin{longtable}{lrrlr}
    \toprule
IRAS name & RA(J2000) & Dec(J2000) & rms\tabnotemark{b}  & Date\tabnotemark{c}   \\  
       & (h:m:s)   & ($^\circ$:$'$:$''$)   & (Jy) &  (yyyy-mm-dd) \\
    \midrule
\endfirsthead

  \tabcaptioncontinued
\toprule
IRAS name & RA(J2000) & Dec(J2000) & rms\tabnotemark{b}  & Date\tabnotemark{c}   \\  
       & (h:m:s)   & ($^\circ$:$'$:$''$)   & (Jy) &  (yyyy-mm-dd) \\
\midrule
\endhead

\bottomrule
\endfoot

\bottomrule
  \tabnotetext{a}{All spectra were centered at $V_{\rm LSR}=0$ km s$^{-1}$ for the $4_2-4_1$ E transition.}
  \tabnotetext{b}{One-sigma rms noise level of the spectra, after correction by elevation-dependent gain and atmospheric opacity}
  \tabnotetext{c}{Date of observation}
\endlastfoot  
 
17021$-$3109 & 17 05 23.42 & $-31$ 13 18.4 & 0.04 & 2010-03-21 \\   
17021$-$3054 & 17 05 24.23 & $-30$ 58 14.4 & 0.04 & 2010-03-21\\
17149$-$3053 & 17 18 11.89 & $-30$ 56 40.8 & 0.04  &  2010-03-03\\
17175$-$2819 & 17 20 42.57 & $-28$ 22 36.8 & 0.03  &  2010-03-03\\
17233$-$2602 & 17 26 28.78 & $-26$ 04 57.8 & 0.03 & 2010-03-21\\
17291$-$2147 & 17 32 10.21 & $-21$ 49 58.9 & 0.014 & 2010-03-21\\
17301$-$2538 & 17 33 14.23 & $-25$ 40 23.5 & 0.03  & 2010-03-03\\
17348$-$2906 & 17 38 04.32 & $-29$ 08 22.7 & 0.03  & 2010-03-03\\
17359$-$2902 & 17 39 08.13 & $-29$ 04 06.1 & 0.03  & 2010-03-03\\
17360$-$2142 & 17 39 05.97 & $-21$ 43 51.8 & 0.021 & 2010-03-01\\
17382$-$2531 & 17 41 20.18 & $-25$ 32 53.3 & 0.03 & 2010-03-03\\
17385$-$2413 & 17 41 38.45 & $-24$ 14 40.9 & 0.020 & 2010-03-01\\
17393$-$2727 & 17 42 32.29 & $-27$ 28 28.2 & 0.03  & 2010-03-03\\
17404$-$2713 & 17 43 37.27 & $-27$ 14 46.4 & 0.03 & 2010-03-03\\
\multicolumn{1}{c}{''}  &  \multicolumn{1}{c}{''} &   \multicolumn{1}{c}{''} & 0.03 &2010-03-21 \\
17479$-$3032 & 17 51 12.59 & $-30$ 33 44.5 & 0.03 & 2010-03-03\\
17482$-$2501 & 17 51 22.55 & $-25$ 01 51.4 & 0.03 & 2010-03-03\\
17487$-$1922 & 17 51 44.78 & $-19$ 23 41.5 & 0.03 & 2010-03-21\\
17506$-$2955 & 17 53 49.42 & $-29$ 55 35.0 & 0.03 & 2010-03-03\\
17516$-$2525 & 17 54 43.45 & $-25$ 26 29.8 & 0.021 & 2010-03-01\\
17540$-$2753 & 17 57 14.06 & $-27$ 54 16.0 & 0.022 & 2010-03-01\\
17548$-$2753 & 17 57 57.94 & $-27$ 53 20.8 & 0.03  & 2010-03-03\\
17550$-$2120 & 17 58 04.30 & $-21$ 21 09.0 & 0.03 & 2010-03-03\\
17550$-$2800 & 17 58 10.72 & $-28$ 00 25.9 & 0.03 & 2010-03-03\\
17552$-$2030 & 17 58 16.37 & $-20$ 30 22.1 & 0.020 & 2010-03-01\\
17560$-$2027 & 17 59 04.61 & $-20$ 27 23.5 & 0.021 & 2010-03-01\\
18011$-$1847 & 18 04 02.81 & $-18$ 47 09.7 & 0.03 & 2010-03-21 \\
18015$-$1352 & 18 04 22.28 & $-13$ 51 49.1 & 0.03 & 2010-03-21\\
18016$-$2743 & 18 04 45.89 & $-27$ 43 11.0 & 0.022 & 2010-03-01 \\
18039$-$1903 & 18 06 53.36 & $-19$ 03 09.3 & 0.025 & 2010-03-01 \\
18049$-$2118 & 18 07 54.93 & $-21$ 18 08.9 & 0.020 & 2010-03-01 \\
18051$-$2415 & 18 08 12.87 & $-24$ 14 35.8 & 0.03 & 2010-03-03\\
18071$-$1727 & 18 10 05.97 & $-17$ 26 35.2 & 0.03 & 2010-03-01 \\
18083$-$2155 & 18 11 18.85 & $-21$ 55 05.1 & 0.03 & 2010-03-01 \\
18087$-$1440 & 18 11 34.66 & $-14$ 39 55.6 & 0.03 & 2010-03-03\\
18105$-$1935 & 18 13 32.33 & $-19$ 35 03.3 & 0.03 & 2010-03-01 \\
18113$-$2503 & 18 14 26.37 & $-25$ 02 55.6 & 0.03  & 2010-03-03\\
\multicolumn{1}{c}{''}  &  \multicolumn{1}{c}{''} &   \multicolumn{1}{c}{''}  & 0.03  & 2010-03-21\\
18135$-$1456 & 18 16 26.16 & $-14$ 55 13.4 & 0.03 & 2010-03-01 \\
18183$-$2538 & 18 21 24.81 & $-25$ 36 35.2 & 0.021 & 2010-03-01 \\
18199$-$1442 & 18 22 50.93 & $-14$ 40 49.4 & 0.024 & 2010-03-01\\
18229$-$1127 & 18 25 45.14 & $-11$ 25 55.7 & 0.03 & 2010-03-21\\
18236$-$0447 & 18 26 20.43 & $-04$ 45 41.8 & 0.025 & 2010-03-01\\
18355$-$0712 & 18 38 15.50 & $-07$ 09 52.3 & 0.023 & 2010-03-01\\
18361$-$1203 & 18 38 58.94 & $-12$ 00 44.3 & 0.024 & 2010-03-01\\
\multicolumn{1}{c}{''}  &  \multicolumn{1}{c}{''} &   \multicolumn{1}{c}{''}   & 0.03 & 2010-03-21\\
18385$+$1350 & 18 40 52.25 & $+13$ 52 53.9 & 0.027 & 2010-03-07\\
18434$-$0042 & 18 46 04.46 & $-00$ 38 55.4 & 0.023 & 2010-03-01\\
18454$+$0001 & 18 48 01.62 & $+00$ 04 48.0 & 0.03  & 2010-03-01\\
18470$+$0015 & 18 49 39.16 & $+00$ 18 52.0 & 0.03  & 2010-03-01\\
18514$+$0019 & 18 53 58.08 & $+00$ 23 25.4 & 0.03  & 2010-03-01\\
18529$+$0210 & 18 55 26.37 & $+02$ 14 48.8 & 0.03  & 2010-03-03\\
\multicolumn{1}{c}{''}  &  \multicolumn{1}{c}{''} &   \multicolumn{1}{c}{''}    & 0.03  & 2010-03-07\\
18580$+$0818 & 19 00 25.31 & $+08$ 22 47.1 & 0.024 & 2010-03-07 \\     
18596$+$0315 & 19 02 06.46 & $+03$ 20 15.1 & 0.025 & 2010-03-07  \\ 
19006$+$1022 & 19 02 59.96 & $+10$ 26 35.1 & 0.024 & 2010-03-21\\
19011$+$1049 & 19 03 30.84 & $+10$ 53 53.3 & 0.024 & 2010-03-07 \\
19015$+$1256 & 19 03 52.75 & $+13$ 01 20.9 & 0.024 & 2010-03-07 \\
19071$+$0857 & 19 09 29.77 & $+09$ 02 23.3 & 0.017 & 2010-03-01\\
\multicolumn{1}{c}{''}  &  \multicolumn{1}{c}{''} &   \multicolumn{1}{c}{''}    & 0.018 & 2010-03-03\\
\multicolumn{1}{c}{''}  &  \multicolumn{1}{c}{''} &   \multicolumn{1}{c}{''}  & 0.023 & 2010-03-07\\
19075$+$0432 & 19 10 00.07 & $+04$ 37 06.2 & 0.3  & 2010-03-03\\
\multicolumn{1}{c}{''}  &  \multicolumn{1}{c}{''} &   \multicolumn{1}{c}{''}  & 0.024 & 2010-03-07\\
19079$-$0315 & 19 10 32.56 & $-03$ 10 15.8 & 0.025 & 2010-03-21 \\
19094$+$1627 & 19 11 44.72 & $+16$ 32 54.0 & 0.024 & 2010-03-07\\
19134$+$2131 & 19 15 35.46 & $+21$ 36 33.2 & 0.024 & 2010-03-07 \\
19178$+$1206 & 19 20 14.24 & $+12$ 12 22.0 & 0.024 & 2010-03-07\\
19181$+$1806 & 19 20 25.28 & $+18$ 11 41.0 & 0.024 & 2010-03-07\\
19190$+$1102 & 19 21 25.37 & $+11$ 08 39.8 & 0.024 & 2010-03-07\\
19193$+$1804 & 19 21 31.65 & $+18$ 10 09.5 & 0.024 & 2010-03-07\\
19315$+$2235 & 19 33 41.75 & $+22$ 42 08.1 & 0.024 & 2010-03-07 \\
19319$+$2214 & 19 34 03.51 & $+22$ 21 13.6 & 0.023 & 2010-03-07\\
19374$+$2359 & 19 39 35.48 & $+24$ 06 24.8 & 0.023 & 2010-03-07 \\
20035$+$3242 & 20 05 29.74 & $+32$ 51 35.1 & 0.022 & 2010-03-07 \\
20042$+$3259 & 20 06 10.73 & $+33$ 07 50.7 & 0.023 & 2010-03-07\\
20214$+$3749 & 20 23 19.29 & $+37$ 58 52.4 & 0.021 & 2010-03-07 \\
20244$+$3509 & 20 26 25.48 & $+35$ 19 14.2 & 0.023 & 2010-03-21 \\
20461$+$3853 & 20 48 04.65 & $+39$ 05 00.7 & 0.024 & 2010-03-21 \\
21525$+$5643 & 21 54 15.14 & $+56$ 57 23.0 & 0.025 & 2010-03-21 \\
21554$+$6204 & 21 56 58.35 & $+62$ 18 43.4 & 0.025 & 2010-03-21 \\
\end{longtable}

\end{document}